\documentclass[twocolumn,10pt,aps,prl,amsmath,amssymb,superscriptaddress]{revtex4-2}
\usepackage[utf8]{inputenc}
\usepackage[english]{babel}
\usepackage{graphicx}
\usepackage{bm}
\usepackage[normalem]{ulem}
\usepackage{xcolor}
\usepackage[colorlinks=true,linkcolor=blue,citecolor=blue,urlcolor=blue]{hyperref}

\newcommand{\dd}{\text d}

\begin{document}
\title{Aging Phase Diagram and Exact Asymptotic Energies of Mixed Spherical Spin Glasses}

\author{Johannes Lang}
\email[]{j.lang@uni-koeln.de}
%\email[]{gu53jup@gmail.com}
\affiliation{Institut f\"ur Theoretische Physik, Universit\"at zu K\"oln,
Z\"ulpicher Stra{\ss}e 77, 50937 Cologne, Germany}

\author{Vincenzo Citro}
\affiliation{DIIN, Universit\`a di Salerno, Via Giovanni Paolo II 132, 84084 Fisciano, Italy}
\affiliation{CNR-Nanotec, Rome Unit, P.~le A.~Moro 5, 00185 Rome, Italy}

\author{Luca Leuzzi}
\affiliation{CNR-Nanotec, Rome Unit, P.~le A.~Moro 5, 00185 Rome, Italy}
\affiliation{Dipartimento di Fisica, Sapienza Universit\`a di Roma, P.~le A.~Moro 5, 00185 Rome, Italy}

\author{Federico Ricci-Tersenghi}
\affiliation{Dipartimento di Fisica, Sapienza Universit\`a di Roma, P.~le A.~Moro 5, 00185 Rome, Italy}
\affiliation{CNR-Nanotec, Rome Unit, P.~le A.~Moro 5, 00185 Rome, Italy}
\affiliation{INFN, Sezione di Roma, P.~le A.~Moro 5, 00185 Rome, Italy}

\date{\today}

\begin{abstract}
%A glass quenched to low temperature never equilibrates: it ages. We derive the aging phase diagram of mixed spherical spin glasses quenched to zero temperature: dynamical Gardner-type transitions separate amorphous states whose effective-temperature hierarchies range from one level to a continuum, each with its exact energy. When the hierarchy is continuous, gradient flow attains the algorithmic bound---relaxation alone is optimal among known polynomial-time algorithms---while all mixtures without pairwise terms fall strictly short: the hierarchy mirrors the landscape's computational complexity.

We determine the aging phase diagram of mixed spherical $(p+s)$-spin glasses quenched from random configurations to zero temperature. Solving the asymptotic dynamical equations of Cugliandolo and Kurchan, we find transitions between aging states with one, two, and continuously many effective temperatures, including phases in which discrete and continuous hierarchies coexist. We obtain analytic expressions for the energies reached asymptotically by the dynamics. In models combining pairwise and higher-order interactions, and sufficiently separated interaction orders ($s>3$), a fully continuous phase exists, where the gradient descent relaxation converges to the algorithmic energy lower bound. Instead, in mixed models with only high-order interactions ($p\ge3$) the energy relaxation remains strictly above it. Numerical integration of the dynamical equations is consistent with the predicted two-step and mixed discrete--continuous aging states.
\end{abstract}
\maketitle

\textit{Introduction.—}
At the mean-field level, glass phases are classified by their pattern of replica symmetry breaking (RSB), which describes the hierarchical organization of the many amorphous states~\cite{Parisi1979a, Parisi1980, SpinGlassBook1987}. Mixed spherical spin glasses, formed by superposing random interactions of different orders, exhibit a particularly rich equilibrium phase diagram. Depending on the composition, one finds one-step, full, and mixed discrete--continuous RSB phases, as well as transitions between distinct amorphous phases~\cite{Crisanti2004, Crisanti2006, Crisanti2007b, Crisanti2011}. The existence of several of these structures has also been established rigorously at zero temperature~\cite{Auffinger2019, Zhou2025}. These results raise the question of whether the non-equilibrium dynamics admits a similarly rich classification.

Following a quench below the dynamical transition, a mean-field glass does not equilibrate. Its relaxation slows down progressively, and correlations depend separately on the observation time and the age of the system~\cite{Bouchaud1998, Cugliandolo2003}. Cugliandolo and Kurchan developed an asymptotic theory of this aging dynamics, first for the pure spherical $p$-spin model and subsequently for the soft spin version of the Sherrington--Kirkpatrick model~\cite{Cugliandolo1993, Cugliandolo1994, Cugliandolo1995}. Different dynamical scales may then be characterized by different effective temperatures~\cite{Cugliandolo1997}. The pure spherical $p$-spin model has a single such temperature throughout the aging regime, whereas the Sherrington--Kirkpatrick model develops a continuous hierarchy. More generally, the asymptotic equations allow several discrete or continuous dynamical scales~\cite{Contucci2021, Kurchan2023, Lang2024b}.

For mixed spherical models, however, the asymptotic dynamics following a quench from a random initial condition has not yet been classified and cannot be inferred from the equilibrium phase diagram. Numerical studies of quenches from finite-temperature states have found regimes with qualitatively different memory of their initial conditions~\cite{Folena2020, Folena2023,Lang2025,Citro2025}. Although these studies concern a different protocol, they demonstrate that mixed interactions can qualitatively alter the late-time dynamics. It remains unknown whether varying the mixture produces transitions between hierarchies of effective temperatures, and what the corresponding transition lines and asymptotic energies are.

Here we determine the non-equilibrium phase diagram of mixed spherical $(p+s)$-spin models quenched from uniformly random configurations to zero temperature. We solve the asymptotic dynamical equations assuming ultrametricity between different timescales and local equilibration on each time scale, classify their stable solutions, and select among competing solutions by their asymptotic energies. For the $(2+s)$ models, increasing the weight of the pairwise interaction produces transitions from a single effective temperature through a mixed discrete--continuous regime to a fully continuous hierarchy. For the $(3+s)$ models, sufficiently separated interaction orders ($s>10$) generate regimes with two discrete temperatures and with discrete and continuous scales, before the single-temperature solution reenters. We determine the transition lines and asymptotic energy in each phase and compare the latter with the algorithmic energy bound. Numerical integration of the full dynamical equations is consistent with both the predicted two-step and mixed discrete--continuous regimes.

\textit{Model and asymptotic theory.—}
The mixed spherical $(p+s)$ model consists of $N$ spins $\sigma_i$ subject to $\sum_i\sigma_i^2=N$. Their random Gaussian energy $\mathcal H(\bm\sigma)$ has zero mean and covariance
\begin{equation}
    \overline{\mathcal H(\bm\sigma)\mathcal H(\bm\sigma')}
    = N f(\bm\sigma\cdot\bm\sigma'/N)\,,
\end{equation}
where $f(q)=\lambda q^p+(1-\lambda)q^s$, with $s>p$ and $0\leq\lambda\leq1$. The limits $\lambda=1$ and $\lambda=0$ recover the pure $p$- and $s$-spin models. Starting from a uniformly random configuration, the spins evolve by gradient flow, corresponding to a quench from $T=\infty$ to $T=0$; the dynamical equations are given in the End Matter.

For $N\to\infty$, the correlation~\footnote{In the $T=0$ gradient flow dynamics, there is no thermal noise, and the angular brackets represent only the average over the random initial condition $\bm\sigma(0)$.}
\begin{equation}
    C(t,t')=\frac{1}{N}\sum_i
    \overline{\langle\sigma_i(t)\sigma_i(t')\rangle}
\end{equation}
and the response $R(t,t')$ obey closed causal equations~\cite{Sompolinsky1981,Sompolinsky1982,Crisanti1993,Cugliandolo1993,Arous2006}. In the aging regime, they are related through
\begin{equation}\label{eq:RXC}
    R(t,t')=X[C(t,t')]\partial_{t'}C(t,t')\,,
\end{equation}
where $1/X[C]$ is the effective temperature at the dynamical scale $C$~\cite{Cugliandolo1997,Herisson2002}.

Following Ref.~\cite{Cugliandolo1994}, we introduce
\begin{equation}\label{eq:FH}
    F[C]=-\!\int_C^1\!\dd c\,X[c],
    \quad
    H[C]=-\!\int_C^1\!\dd c\,f''(c)X[c].
\end{equation}
Time-reparametrization invariance implies a composition law for correlations at three widely separated times $t_{\min}\!<\!t_{\text{int}}\!<\!t_{\max}$: $C(t_{\max},t_{\min})=g(C(t_{\max},t_\text{int}),C(t_\text{int},t_{\min}))$.
We denote its generalized inverse by $\bar g$, defined through
\begin{equation}\label{eq:gbar}
    C(t_{\max},t_\text{int})
    =
    \bar g\big(
        C(t_{\text{int}},t_{\min}),
        C(t_{\max},t_{\min})
    \big)\,.
\end{equation}
The asymptotic response equation then takes the reparametrization-invariant form
\begin{align}
    0=&-\sqrt{f''(1)}\,F[C]
    +\frac{H[C]}{\sqrt{f''(1)}}+\nonumber\\
    &+\int_C^1\dd c\,f''(c)X[c]F[\bar g(c,C)]\,.
    \label{eq:R}
\end{align}
Under local equilibration, the asymptotic correlation equation yields the same endpoint conditions as Eq.~\eqref{eq:R}, as shown in the End Matter.

A discrete dynamical scale is characterized by a constant $X[C]$, while a continuous hierarchy is described by a continuously varying $X[C]$. The scales are bounded by endpoints $a_n^*$ satisfying
\begin{equation}
    g(a_n^*,a_n^*)=a_n^*\,, \quad \bar g(a_n^*,a_n^*)=a_n^*\,,\quad a_{n+1}^* \leq a_n^*\,.
\end{equation}
For a discrete scale $a_{n+1}^*<a_n^*$, whereas the $a_n^*$ are dense in a continuous hierarchy.
In analogy with the Parisi classification~\cite{Parisi1979a, Parisi1980, SpinGlassBook1987, Crisanti2007, Kurchan2023, Lang2024b}, we call a solution $k$-RSB if $X[C]$ has $k$ plateaus, FRSB (full RSB) if it varies continuously, and $k$-FRSB if $k$ plateaus coexist with a continuous branch.

% \begin{table}[!b]
% \caption{effective inverse temperature $X[C]$ in the five aging states, with $X_{\text{\tiny FRSB}}(C)=f'''(C)/[2f''(C)^{3/2}]$. Scales satisfy $0\leq a_2^*<a_1^*\leq1$; in all phases $X'[C]\geq0$.}
% \label{tab:states}
% \begin{ruledtabular}
% \begin{tabular}{ll}
% State & $X[C]$ \\
% \hline
% 1-RSB & $X_1$ \\
% 2-RSB & $X_2$ for $C<a_1^*$;\quad $X_1$ for $C>a_1^*$ \\
% 1-FRSB ($p=2$) & $X_{\text{\tiny FRSB}}(C)$ for $C\leq a_1^*$;\quad $X_1$ for $C>a_1^*$ \\
% 2-FRSB & $X_2$ for $C<a_2^*$;\quad $X_{\text{\tiny FRSB}}(C)$ on $(a_2^*,a_1^*)$;\\
%  & $X_1$ for $C>a_1^*$ \\
% 1-FRSB ($p\geq3$) & $X_1$ for $C<a_1^*$;\quad $X_{\text{\tiny FRSB}}(C)$ for $C\geq a_1^*$ \\
% FRSB & $X_{\text{\tiny FRSB}}(C)$ \\
% \end{tabular}
% \end{ruledtabular}
% \end{table}

The appearance of a new scale follows directly from Eq.~\eqref{eq:R}. Evaluating it and its derivative at an endpoint $a_n^*$ eliminates $\bar g$ and gives
\begin{equation}\label{eq:solvability}
    \begin{split}
        F[a_n^*] &= \frac{1}{\sqrt{f''(1)}} - \frac{1}{\sqrt{f''(a_n^*)}}\,,\\
        H[a_n^*] &= \sqrt{f''(a_n^*)} - \sqrt{f''(1)}\,.
    \end{split}
\end{equation}
These conditions apply at every scale endpoint and determine both the appearance and location of additional scales, while the value of $X[C]$ on the lowest scale follows from marginality.

\begin{figure}[t]
\centering
\includegraphics[width=0.9\columnwidth]{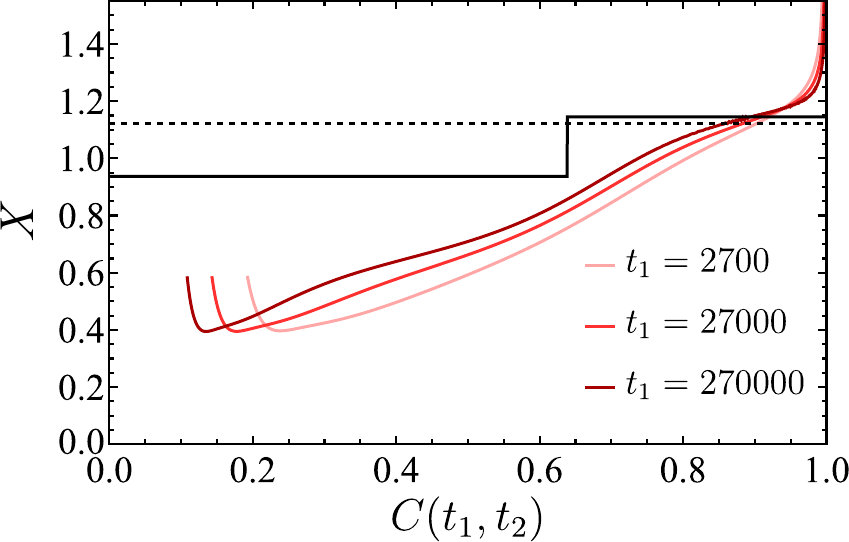}\\
\hspace{0.25cm}\\
\includegraphics[width=0.9\columnwidth]{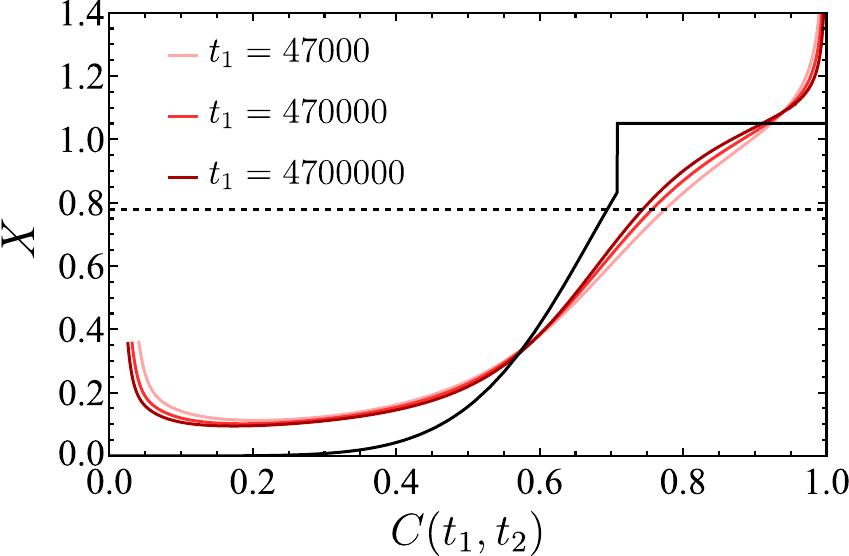}
\caption{Test of the asymptotic solutions against numerical integration of the full dynamical equations~\cite{Lang2026}. Colored curves show $X(t_1,t_2)=R(t_1,t_2)/\partial_{t_2}C(t_1,t_2)$ versus $C(t_1,t_2)$ for increasing $t_1$. Black lines are the parameter-free asymptotic predictions; dashed lines show the marginal 1-RSB values $X_\text{mg}$. \textbf{Top}: $(3+15)$ model at $\lambda=0.8$ in the 2-RSB phase, with $X_2\simeq0.94$ for $C\lesssim0.64$, $X_1\simeq1.15$ for $C\gtrsim0.64$, and $X_\text{mg}\simeq1.12$. \textbf{Bottom}: $(2+9)$ model at $\lambda=0.9$ in the 1-FRSB phase, with a high-correlation plateau $X_1\simeq1.05$ and $X_\text{mg}\simeq0.78$. In both panels, the finite-time results drift toward the predicted asymptotic form.}
\label{fig:X}
\end{figure}

The construction assumes ultrametricity and local equilibration, but not weak ergodicity breaking. The explicit solution for $\bar g$ shows that all solutions obtained here are weak glasses: correlations with any fixed earlier time eventually decay to zero~\cite{companion}. When several stable asymptotic solutions coexist, we assume that gradient flow selects the one with the highest energy, corresponding to the first stable solution reached as the energy decreases. The selected solutions satisfy $X'[C]\geq0$. We test this selection criterion against the full dynamical equations in Fig.~\ref{fig:X}.

Together with the endpoint conditions, the energy criterion determines the non-equilibrium phase diagram. We now apply this construction to the $(2+s)$ and $(3+s)$ models, which exhibit distinct sequences of asymptotic phases.

\begin{figure}[t]
\centering
\includegraphics[width=0.9\columnwidth]{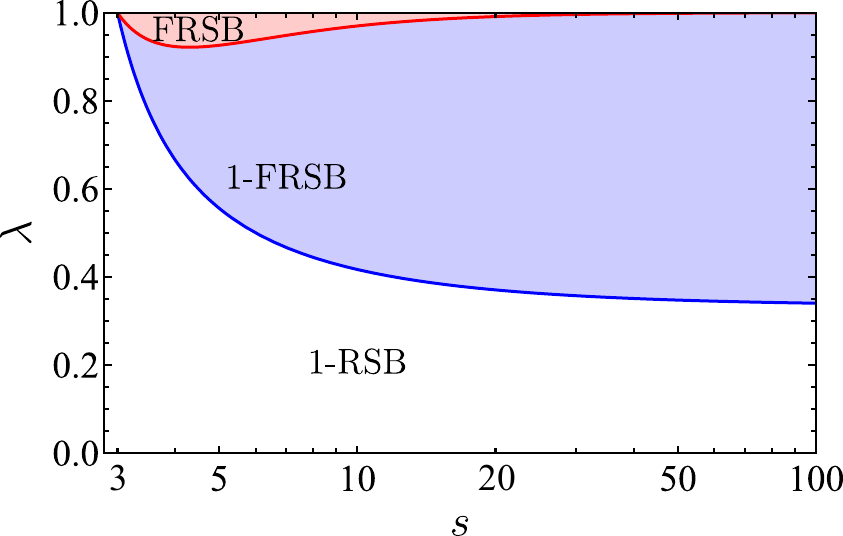}
\caption{Non-equilibrium phase diagram of the $(2+s)$ models quenched from infinite to zero temperature, with $s$ shown on a logarithmic scale. The blue and red lines are the 1-RSB--1-FRSB and 1-FRSB--FRSB transitions given by Eqs.~\eqref{eq:lambda2s} and \eqref{eq:lambda1FF}. In the FRSB phase, gradient flow reaches the algorithmic bound $E_{\text{alg}}$.}
\label{fig:PD2s}
\end{figure}

\textit{The $(2+s)$ models.—}
For every $s>3$, as in the thermodynamic counterpart \cite{Crisanti2004,Crisanti2006}, increasing $\lambda$ from the pure $s$-spin limit produces the sequence 1-RSB, 1-FRSB, and FRSB shown in Fig.~\ref{fig:PD2s}. The 1-RSB solution has $X[C]=X_\text{mg}$, with marginal stability fixing
\begin{equation}
    X_\text{mg}=\frac{\sqrt{f''(1)}}{f'(1)}
        -\frac{1}{\sqrt{f''(1)}},
\end{equation}
which generalizes the threshold solution of the pure model~\cite{Cugliandolo1993}. A second scale becomes possible when Eqs.~\eqref{eq:solvability}, evaluated with $X[C]=X_\text{mg}$, first admit an endpoint $a_1^*\in[0,1)$. For $p=2$, this instability occurs at $a_1^*=0$ and gives the analytic transition line
\begin{equation}\label{eq:lambda2s}
    \lambda_{\text{1-1F}}=\frac{s}{3(s-2)}.
\end{equation}

Beyond this line, a continuous hierarchy develops for $C\leq a_1^*$, while a discrete scale remains for $C>a_1^*$. On the continuous branch, Eq.~\eqref{eq:R} reduces to a local identity and fixes
\begin{equation}\label{eq:X1FRSB}
    X[C]=
    \begin{cases}
        X_1 & C>a_1^*\,,\\[2pt]
        X_{\text{\tiny FRSB}}(C)
        \equiv\dfrac{f'''(C)}{2f''(C)^{3/2}}
        & C\leq a_1^*\,.
    \end{cases}
\end{equation}
The endpoint $a_1^*$ and plateau value $X_1$ are determined by Eqs.~\eqref{eq:solvability}, which already incorporate marginality through their derivation from the asymptotic equation of motion~\eqref{eq:R}. This 1-FRSB phase (see Table~\ref{tab:states}) is the dynamical counterpart of the mixed discrete--continuous phase in the equilibrium diagram~\cite{Crisanti2004, Crisanti2006}.

With increasing $\lambda$, the plateau contracts until $a_1^*\to1$ on the second analytical line
\begin{equation}\label{eq:lambda1FF}
    \lambda_{\text{1F-F}}
    =\frac{s^2(s-1)}{(s-2)(s^2+s+6)}\,.
\end{equation}
Above this line, $X[C]=X_{\text{\tiny FRSB}}(C)$ for $0\leq C\leq1$, and the asymptotic dynamics has a continuum of effective temperatures. The endpoint conditions admit no further instability. Both transition lines terminate at $\lambda=1$ for $s=3$, so the $(2+3)$ model is 1-RSB for every $\lambda<1$.

\begin{figure}[t]
\centering
\includegraphics[width=0.9\columnwidth]{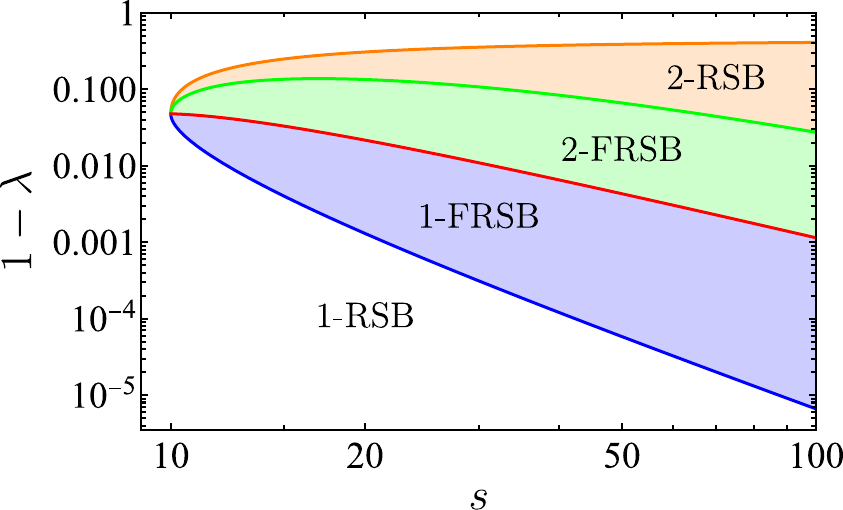}
\caption{Non-equilibrium phase diagram of the $(3+s)$ models, plotted as $1-\lambda$ versus $s$ on logarithmic axes. Increasing $\lambda$ (from top to bottom) yields the sequence 1-RSB, 2-RSB, 2-FRSB, 1-FRSB, and reentrant 1-RSB. The critical lines $\lambda_{\text{1-2}}$ (orange), $\lambda_{\text{2-2F}}$ (green) and $\lambda_{\text{2F-1F}}$ (red) are determined numerically, while $\lambda_{\text{1F-1}}$ (blue) is given analytically by Eq.~\eqref{eq:lambda1F1}. All four lines meet at $s=10$ and $1-\lambda=1/21$. For $s<10$, the asymptotic solution is 1-RSB for every $\lambda\in[0,1]$.}
\label{fig:PD3s}
\end{figure}

\textit{The $(3+s)$ models.—}
 For $s <10$ the only off-equilibrium dynamical phase reached in the zero-temperature quench is 1-RSB, for any $\lambda$. For {$s>10$}, increasing $\lambda$ from the pure $s$-spin limit produces the sequence 1-RSB, 2-RSB, 2-FRSB, 1-FRSB, and the reentrant 1-RSB shown in Fig.~\ref{fig:PD3s}, in complete analogy to the static solutions of the model \cite{Crisanti2007b,Crisanti2011}. The first transition occurs when Eqs.~\eqref{eq:solvability}, evaluated for $X[C]=X_\text{mg}$, admit an endpoint at finite $a_1^*$. The additional plateau has finite width at its onset, while its height relative to $X_\text{mg}$ vanishes. We therefore identify this continuous bifurcation of the RSB hierarchy as a fragmentation transition~\cite{Crisanti2007b}. It produces the 2-RSB phase and defines the line $\lambda_{\text{1-2}}$.

At $\lambda_{\text{2-2F}}$, the curve $X_{\text{\tiny FRSB}}(C)$ meets the two-plateau solution at its breakpoint $(a_1^*,X_2)$. Beyond this line, a continuous branch exists between the two plateaus, and the maximum-energy solution is the 2-FRSB one. At $\lambda_{\text{2F-1F}}$, the higher plateau contracts to zero width ($a_1^*\to1$). The resulting 1-FRSB solution has the reversed structure listed in Table~\ref{tab:states}: a plateau at low correlations and a continuous branch extending to $C=1$. This arrangement is opposite to the $(2+s)$ case: for $p\geq3$, $X'_{\text{\tiny FRSB}}(C)$ becomes negative at small $C$, so the continuous branch cannot extend to $C=0$. Marginality and maximization of the asymptotic energy determine the endpoint and plateau value, with energy maximization enforcing continuity at their junction.

The 1-RSB phase reenters when the continuous branch shrinks to zero
and the 1-FRSB solution merges with the marginal 1-RSB solution,
$X_{\mathrm{FRSB}}(1)=X_{\mathrm{mg}}$. The corresponding line
$\lambda_{\mathrm{1F-1}}$ is obtained analytically for general $p$ in
Eq.~\eqref{eq:lambda1F1} of the End Matter. For $p=3$, a physical
solution exists only for $s\geq10$. At $s=10$, all four phases meet at
$\lambda=20/21$ (Fig.~\ref{fig:PD3s}). In each solution, the upper
endpoint of the lowest discrete scale approaches $a_1^*=1$, compressing
all additional scale structure into $C=1$. The solutions therefore
coincide with the marginal 1-RSB solution at this multicritical point.

Except for $\lambda_{\text{1F-1}}$, the transition lines are obtained numerically by solving the endpoint, marginality, and energy conditions. This finite algebraic problem can be evaluated to arbitrary precision.

\begin{table}[t]
\caption{Effective inverse temperature $X[C]$ in the asymptotic phases. The endpoints satisfy $0\leq a_2^*<a_1^*\leq1$.}
\label{tab:states}
\centering
\begin{tabular}{@{}ll@{}}
\hline\hline
Phase & $X[C]$ \\
\hline
1-RSB
& $X_{\mathrm{mg}}$ \\[2pt]
2-RSB
& $X_2$ for $C<a_1^*$;\quad $X_1$ for $C>a_1^*$ \\[2pt]
1-FRSB ($p=2$)
& $X_{\text{\tiny FRSB}}(C)$ for $C\leq a_1^*$;\quad
  $X_1$ for $C>a_1^*$ \\[2pt]
2-FRSB
& $X_2$ for $C<a_2^*$;\quad
  $X_{\text{\tiny FRSB}}(C)$ for $a_2^*<C<a_1^*$; \\
& $X_1$ for $C>a_1^*$ \\[2pt]
1-FRSB ($p\geq3$)
& $X_2$ for $C<a_2^*$;\quad
  $X_{\text{\tiny FRSB}}(C)$ for $C\geq a_2^*$ \\[2pt]
FRSB
& $X_{\text{\tiny FRSB}}(C)$ \\
\hline\hline
\end{tabular}
\end{table}

\begin{figure}[t]
\centering
\includegraphics[width=0.9\columnwidth]{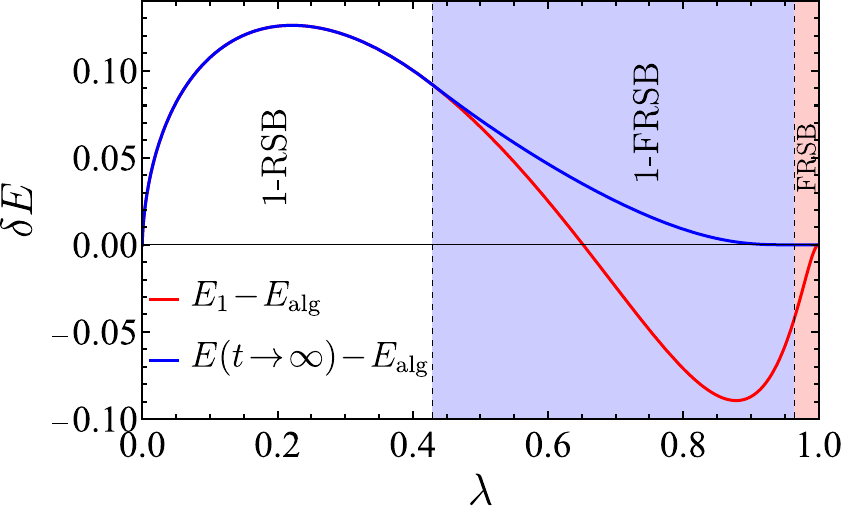}\\
\hspace{0.25cm}\\
\includegraphics[width=0.9\columnwidth]{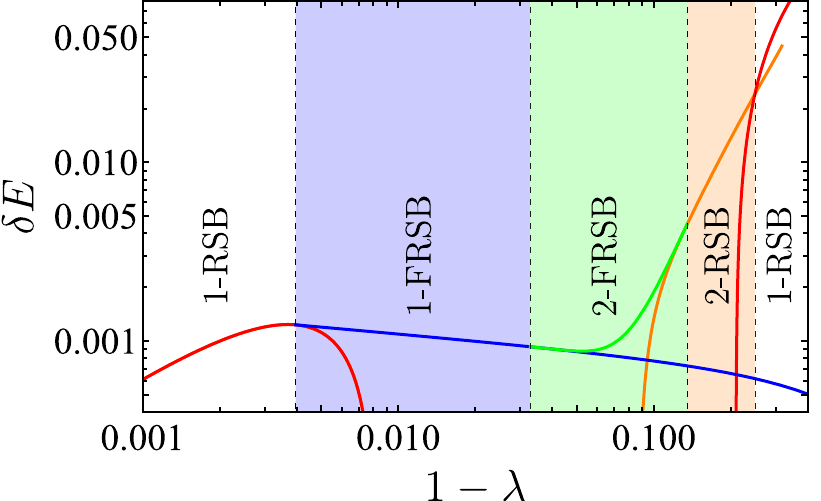}
\caption{Asymptotic energies relative to the algorithmic bound $E_{\text{alg}}$. \textbf{Top}: physical energy $E_\infty-E_{\text{alg}}$ (blue) and continued marginal 1-RSB energy $E_1-E_{\text{alg}}$ (red) for the $(2+9)$ model. The dashed lines mark $\lambda_{\text{1-1F}}=3/7$ and $\lambda_{\text{1F-F}}\simeq0.964$. The 1-FRSB solution solves the paradox of having $E_1<E_\text{alg}$ for $\lambda \gtrsim 0.65$. In the FRSB phase, $E_\infty=E_{\text{alg}}$. \textbf{Bottom}: energies of the 1-RSB (red), 1-FRSB (blue), 2-FRSB (green), and 2-RSB (orange) solutions of the $(3+15)$ model, plotted against $1-\lambda$. Each branch is shown over the domain in which the corresponding asymptotic solution exists. The physical energy is their upper envelope, which remains strictly above $E_{\text{alg}}$ for any $\lambda$ value. The dashed lines mark the transitions at $\lambda\simeq0.752$, $0.865$, $0.967$, and $0.996$.}
\label{fig:energy}
\end{figure}

\textit{Exact asymptotic energies.—}
For a given asymptotic solution $X[C]$, the energy density reached by gradient flow is given by the following expression derived in the End Matter:
\begin{equation}\label{eq:Einf}
    E_\infty\equiv E(t\to\infty)
    =-\frac{f'(1)}{\sqrt{f''(1)}}
     -\int_0^1\dd c\,f'(c)X[c]\,.
\end{equation}
When several stable solutions coexist, the maximum-energy criterion selects the physical one. Accordingly, $E_\infty$ is the upper envelope of the corresponding energies in Fig.~\ref{fig:energy}. Its nonanalyticity distinguishes the fragmentation transition from the remaining phase boundaries: $\partial_\lambda E_\infty$ is discontinuous at $\lambda_{\text{1-2}}$, whereas both $E_\infty$ and its first derivative are continuous across all other transitions.

To assess the efficiency of gradient flow, we compare $E_\infty$ with the algorithmic bound, that is, the lowest energy reachable by known algorithms in a time scaling polynomially in $N$~\cite{Subag2021, ElAlaoui2021, ElAlaoui2020},
\begin{equation}\label{eq:Ealg}
    E_{\text{alg}}
    =-\int_0^1\dd c\,\sqrt{f''(c)}\,.
\end{equation}
This bound has been proven for any stable algorithm~\footnote{An algorithm is defined as stable if its output depends Lipschitz-continuously on the parameters entering the Hamiltonian to be minimized.} minimizing an even mixture~\cite{HuangSellke2025}, i.e.\ one containing only even powers of $q$, and the result is conjectured to extend to general mixtures~\cite{Gamarnik2021, ElAlaoui2020}. The class of stable algorithms includes gradient flow and Langevin dynamics run for any bounded time.

The comparison also clarifies why the marginal 1-RSB solution cannot describe the entire phase diagram. Let $E_1$ denote its energy when continued away from the pure-model limits. For some mixtures, $E_1$ falls below $E_{\text{alg}}$ (see, e.g., $\lambda\gtrsim 0.65$ in the top panel of Fig.~\ref{fig:energy}), reproducing the previously observed breakdown of the conventional threshold construction~\cite{Folena2020, Folena2021, Folena2023, Kent-Dobias2024b}. The additional scales found here replace the unstable 1-RSB solution with a richer 1-FRSB solution, whose asymptotic energy is above $E_1$. The resulting $E_\infty$ never falls below the algorithmic bound in the entire phase diagram (see Fig.~\ref{fig:energy}).

In the FRSB phase, gradient flow saturates the bound,
\begin{equation}
    E_\infty=E_{\text{alg}}\,.
\end{equation}
This happens for $p=2$, $s>3$ and $\lambda\geq\lambda_{\text{1F-F}}$. Thus, throughout this phase, gradient flow is optimal among known polynomial-time algorithms, extending the corresponding result for the pure $p=2$ model~\cite{Cugliandolo1995b}. For mixed $(p+s)$ models with $p\geq3$, by contrast, the FRSB phase is absent and $E_\infty$ remains strictly above $E_{\text{alg}}$ for every $0<\lambda<1$. Equality is recovered at the pure-model endpoints~\cite{Sellke2024, Subag2021}.

\textit{Numerical test.—}
Ultrametricity, local equilibration, and maximum-energy selection lead to parameter-free predictions for $X[C]$, which can be tested by integrating the full dynamical mean-field equations at finite times. Using the method of Ref.~\cite{Lang2026}, we consider one mixture in the 2-RSB phase and one in the 1-FRSB phase.

After verifying that the extrapolation of the energy $E(t)$ is consistent with the asymptotic value $E_\infty$ shown in Fig.~\ref{fig:energy}, we report in Fig.~\ref{fig:X} the parametric plot of $X(t_1,t_2)$ versus $C(t_1,t_2)$ that converges to $X[C]$, as $t_1$ increases.
The upper panel reports data for the $(3+15)$ model at $\lambda=0.8$: the curves slowly evolve
toward the predicted two-step form, with two plateaus separated at $C\simeq0.64$. The lower panel shows data for the $(2+9)$ model at $\lambda=0.9$: the numerical results approach the predicted 1-FRSB solution, consisting of a continuous branch at lower correlations and a plateau at $X_1\simeq1.05$.

In both cases, the approach to the asymptotic prediction is systematic but very slow, particularly at small correlations. Compared with the pure models, the dynamics is substantially slower. This is consistent with the additional hierarchy of time scales associated with the more complex RSB structure, although a quantitative analysis of the finite-time evolution is beyond the scope of the present work.

\textit{Discussion.—}
Figures~\ref{fig:PD2s} and~\ref{fig:PD3s} provide the asymptotic phase diagrams of mixed spherical spin glasses quenched from random configurations to zero temperature. Varying the interaction mixture changes the hierarchy of dynamical scales selected by the same protocol, producing transitions between one-step, multi-step, and continuous RSB regimes. Their boundaries follow from the asymptotic equations, in closed form where possible and otherwise by numerical solution to arbitrary precision.

The transitions fall into two classes. At the boundary between 1-RSB and 2-RSB phases, the energies of the two phases cross (as the free energy does at a first-order transition~\cite{Crisanti2007b}) and the asymptotic energy $E_\infty$ shows a cusp; here, however, the two solutions coincide at the crossing ($X_1=X_2=X_\text{mg}$), so that no distinct states coexist. Instead, at the remaining boundaries, the transitions are continuous and Gardner-like: they always involve a continuous RSB sector, and both $E_\infty$ and its first derivative are continuous. Together, these transitions constitute the non-equilibrium counterparts of the amorphous--amorphous and Gardner transitions in equilibrium glasses~\cite{Charbonneau2014, Berthier2019a, Parisi2020}.

Unlike conventional Gardner protocols, which follow an equilibrated glass as temperature, density, or another external parameter is varied~\cite{Rainone2015, Berthier2016, Hammond2020}, the present phase diagrams classify the asymptotic dynamics reached from random initial conditions under the same quench. The resulting effective-temperature hierarchies are, in principle, observable through parametric measurements of correlations and responses~\cite{Cugliandolo1997, Herisson2002, Sciortino2001}. The nonequilibrium quench diagram and the equilibrium phase diagram at zero temperature contain the same RSB phases in the same order as $\lambda$ is increased~\cite{Crisanti2004, Crisanti2006, Crisanti2007b, Crisanti2011, Auffinger2019, Auffinger2022, Zhou2025}, but their transition lines and multicritical structures differ. In particular, all intermediate phases of the non-equilibrium $(3+s)$ diagram meet at $s=10$. 
%\luca{Since these intermediate phases separate the low-$\lambda$ 1-RSB solution from the reentrant one over the whole range, the non-equilibrium diagram also inherits the \emph{avoided} first-order structure of the statics,
%cf. point C of Fig.~4 of Ref.~\cite{Crisanti2007b}.}

The exact asymptotic energies provide a complementary connection to optimization. In the FRSB phase of the $(2+s)$ models, gradient flow reaches $E_{\mathrm{alg}}$ throughout a finite parameter region and is optimal among known polynomial-time algorithms. For mixed models with $p\geq3$, its energy remains strictly above the bound, and $E_\infty-E_{\mathrm{alg}}$ quantifies the improvement available to more elaborate algorithms~\cite{ElAlaoui2020, ElAlaoui2021, Montanari2025}. The numerical approach toward the predicted 2-RSB and 1-FRSB solutions supports the asymptotic construction despite the exceptionally slow finite-time evolution. Extensions to finite-temperature initial conditions and non-gradient dynamics remain open.

\textit{Acknowledgments.—}
J.L. thanks Konstantin Weisenberger for useful discussions. The work of J.L. was supported by the Deutsche Forschungsgemeinschaft (DFG, German Research Foundation) under Germany’s Excellence Strategy Cluster of Excellence Matter and Light for Quantum Computing (ML4Q) EXC 2004/1 390534769, and by the DFG Collaborative Research Center (CRC) 1238 Project No. 277146847. 

\textit{Data availability.—}
The results can be reproduced by integrating the dynamical equations
given in the End Matter using the method of Ref.~\cite{Lang2026}.
The full time-evolution data are not publicly archived because of
their size. The processed data underlying the figures are available
from the corresponding author upon reasonable request.

\bibliography{bibliography}

\clearpage

\clearpage

\appendix*
\setcounter{equation}{0}

\section*{End Matter}

\textit{Dynamical equations.—}
The system is prepared at the initial time $t=0$ in a uniformly random configuration, corresponding to equilibrium at $T=\infty$, and then evolves by gradient flow,
\begin{equation}\label{eq:GF}
    \partial_t\sigma_i(t)
    =-\mu(t)\sigma_i(t)
     -\frac{\partial\mathcal H(\bm\sigma(t))}
            {\partial\sigma_i},
\end{equation}
where $\mu(t)$ enforces the spherical constraint. The response to a
field $h_i(t')$ conjugate to $\sigma_i(t')$ is
\begin{equation}
    R(t,t')
    =\frac{1}{N}\sum_i
      \overline{
      \left.
      \frac{\delta\langle\sigma_i(t)\rangle}
           {\delta h_i(t')}
      \right|_{h=0}}.
\end{equation}
For $N\to\infty$, the correlation and response obey the closed causal
equations~\cite{Crisanti1993,Cugliandolo1993,Arous2006}
\begin{widetext}
\begin{align}\label{eq:EOM_Glass}
\begin{split}
    \partial_t C(t,t')
    &=-\mu(t)C(t,t')
      +\int_0^t\dd s\,
       f''(C(t,s))R(t,s)C(s,t')
      +\int_0^{t'}\dd s\,
       f'(C(t,s))R(t',s),\\
    \partial_t R(t,t')
    &=-\mu(t)R(t,t')
      +\int_{t'}^t\dd s\,
       f''(C(t,s))R(t,s)R(s,t'),\\
    \mu(t)
    &=\int_0^t\dd s\,
      \left[
      f''(C(t,s))C(s,t)+f'(C(t,s))
      \right]R(t,s).
\end{split}
\end{align}
\end{widetext}
These are also the equations integrated numerically in
Fig.~\ref{fig:X}.

Following Ref.~\cite{Cugliandolo1993}, we separate the time-translation-invariant dynamics close to the time diagonal ($t\approx t'$) from the aging contribution. The stationary response is
\begin{equation}\label{eq:RTTI}
    R_{\mathrm{TTI}}(\tau)
    =
    e^{-2\sqrt{f''(1)}\,\tau}\,
    {}_0F_1\!\left(;2;f''(1)\tau^2\right).
\end{equation}
Matching the stationary and aging regimes fixes the asymptotic
Lagrange multiplier to
\begin{equation}\label{eq:muinf}
    \mu_\infty
    \equiv\lim_{t\to\infty}\mu(t)
    =2\sqrt{f''(1)}.
\end{equation}
This is the marginality condition used in the main text. After
subtracting the stationary contribution, the aging components, whose
label we suppress below, satisfy
$\partial_tC=\partial_tR=0$ to leading order. The response equation
then becomes
\begin{multline}\label{eq:Raging}
    0=
    \left[
    -\sqrt{f''(1)}
    +\frac{f''(C(t,t'))}{\sqrt{f''(1)}}
    \right]R(t,t')
    \\
    +\int_{t'}^t\dd s\,
    f''(C(t,s))R(t,s)R(s,t').
\end{multline}

\textit{Asymptotic response equation.—}
For three asymptotically separated times, the triangle relation defines
a correlation-composition law and its generalized inverse $\bar g$,
as in Eq.~\eqref{eq:gbar}. Within a dynamical scale, $\bar g$ describes
the nontrivial composition of correlations. Correlations belonging to
different scales compose ultrametrically,
\begin{equation}\label{eq:gbar-cross}
    \bar g(C,C')=\min(C,C')\,.
\end{equation}
For the ordered arguments $C'\leq C$ occurring below, this reduces to
$\bar g(C,C')=C'$.

The definitions in Eq.~\eqref{eq:FH} imply
\begin{align}
\begin{split}
    \frac{\dd}{\dd t'}F[C(t,t')]&=R(t,t'),\\
    \frac{\dd}{\dd t'}H[C(t,t')]
    &=f''(C(t,t'))R(t,t'),
\end{split}
\end{align}
together with
\begin{equation}
    F'[C]=X[C],\qquad
    H'[C]=f''(C)X[C].
\end{equation}
Using these relations and the triangle composition law, integration
of Eq.~\eqref{eq:Raging} over $t'$ gives
\begin{multline}\label{eq:REM}
    0=-\sqrt{f''(1)}\,F[C]
      +\frac{H[C]}{\sqrt{f''(1)}}\\
      +\int_C^1\dd c\,
       f''(c)X[c]F[\bar g(c,C)],
\end{multline}
which is Eq.~\eqref{eq:R} of the main text.

Let $a_1^*$ be the upper endpoint of a dynamical scale. For
$c\geq a_1^*$, the two correlations belong to different scales, and
Eq.~\eqref{eq:gbar-cross} gives
$\bar g(c,a_1^*)=a_1^*$. Evaluating Eq.~\eqref{eq:REM} at this endpoint
therefore yields
\begin{equation}\label{eq:Rendpoint}
\begin{split}
    0={}&-\sqrt{f''(1)}\,F[a_1^*]
      +\frac{H[a_1^*]}{\sqrt{f''(1)}}-H[a_1^*]F[a_1^*].
\end{split}
\end{equation}
Differentiating Eq.~\eqref{eq:REM} with respect to $C$ before taking
the endpoint gives
\begin{equation}\label{eq:dREM}
\begin{split}
0={}&-\sqrt{f''(1)}X[C]
 +\frac{f''(C)}{\sqrt{f''(1)}}X[C]\\
&-f''(C)X[C]F[\bar g(C,C)]\\
&+\int_C^1\dd c\,
 f''(c)X[c]\,
 \partial_C F[\bar g(c,C)].
\end{split}
\end{equation}
At $C=a_1^*$, local equilibration gives
\begin{equation}
    \partial_C F[\bar g(c,C)]\big|_{C=a_1^*}
    =X[a_1^*],\qquad c>a_1^*,
\end{equation}
while $\bar g(a_1^*,a_1^*)=a_1^*$. Using
\begin{equation}
    \int_{a_1^*}^1\dd c\,f''(c)X[c]
    =-H[a_1^*],
\end{equation}
Eq.~\eqref{eq:dREM} becomes, after division by
$X[a_1^*]>0$,
\begin{equation}\label{eq:dRendpoint}
    0=-\sqrt{f''(1)}
      +\frac{f''(a_1^*)}{\sqrt{f''(1)}}
      -H[a_1^*]
      -f''(a_1^*)F[a_1^*].
\end{equation}
Solving Eqs.~\eqref{eq:Rendpoint} and
\eqref{eq:dRendpoint} on the branch $F\leq0$ gives
Eqs.~\eqref{eq:solvability} of the main text. The derivation applies
both to a discrete scale $a_2^*<a_1^*$ and, in the limit
$a_2^*\to a_1^*$, to an infinitesimal scale within a continuous
hierarchy.

If $X[C]=X_1$ for $C>a_1^*$, then
\begin{equation}
    F[a_1^*]=(a_1^*-1)X_1,\;
    H[a_1^*]=[f'(a_1^*)-f'(1)]X_1,
\end{equation}
so that Eqs.~\eqref{eq:solvability} determine
$(a_1^*,X_1)$. At the instability of the 1-RSB solution, the plateau
value is
\begin{equation}
    X_{\mathrm{mg}}
    =\frac{\sqrt{f''(1)}}{f'(1)}
     -\frac{1}{\sqrt{f''(1)}}.
\end{equation}
For the $(2+s)$ models, the new scale first appears at $a_1^*=0$.
Eqs.~\eqref{eq:solvability} then reduce to
\begin{equation}
    f''(0)f''(1)=f'(1)^2,
\end{equation}
which gives Eq.~\eqref{eq:lambda2s}.

\textit{Asymptotic correlation equation.—}
Within a dynamical scale $a_2^*<C<a_1^*$, the correlation equation
becomes
\begin{equation}\label{eq:C}
\begin{split}
0={}&-\sqrt{f''(1)}\,C
 +\frac{f'(C)}{\sqrt{f''(1)}}
 -\int_0^C\dd c'\,
 f''(c')F[\bar g(C,c')]\\
&+\int_C^1\dd c'\,
 f''(c')X[c']\,\bar g(c',C)\\
&+\int_0^C\dd c'\,
 f''(c')X[c']\,\bar g(C,c')\,.
\end{split}
\end{equation}
Local equilibration implies that $X[C]$ is constant within this
scale. Differentiating Eq.~\eqref{eq:C} with respect to $C$ and
evaluating it at $C=a_1^*$ gives
\begin{equation}\label{eq:Cendpoint}
    0=-\sqrt{f''(1)}
      +\frac{f''(a_1^*)}{\sqrt{f''(1)}}
      -H[a_1^*]
      -f''(a_1^*)F[a_1^*].
\end{equation}
This coincides with Eq.~\eqref{eq:dRendpoint}. The correlation equation
therefore imposes no additional condition on the instability of a
locally equilibrated solution.

\textit{Continuous hierarchy.—}
On an interval where $X[C]$ varies continuously, every value of $C$
is the endpoint of an infinitesimal scale. Equation~\eqref{eq:Cendpoint}
therefore holds throughout the interval:
\begin{equation}\label{eq:local}
    0=-\sqrt{f''(1)}
      +\frac{f''(C)}{\sqrt{f''(1)}}
      -H[C]-f''(C)F[C].
\end{equation}
Together with Eq.~\eqref{eq:FH}, this gives
\begin{equation}
    F[C]
    =\frac{1}{\sqrt{f''(1)}}
     -\frac{1}{\sqrt{f''(C)}},
\end{equation}
and hence
\begin{equation}
    X[C]=X_{\text{\tiny FRSB}}(C)
    =\frac{f'''(C)}{2f''(C)^{3/2}},
\end{equation}
as stated in Eq.~\eqref{eq:X1FRSB}.

\textit{Analytic reentrant line.—}
For the $(p+s)$ models with $p\geq3$, the 1-FRSB solution merges with
the reentrant 1-RSB solution when the endpoint of its continuous branch
reaches $C=1$. At this point,
\begin{equation}\label{eq:mergecondition}
    X_{\mathrm{FRSB}}(1)=X_{\mathrm{mg}},
\end{equation}
or, equivalently,
\begin{equation}\label{eq:mergecondition-f}
    f'(1)f'''(1)
    =2f''(1)\bigl[f''(1)-f'(1)\bigr].
\end{equation}
Substituting $f(q)=\lambda q^p+(1-\lambda)q^s$ into Eq.~\eqref{eq:mergecondition-f} gives a quadratic equation for $\lambda$. Selecting the physical root yields
\begin{equation}\label{eq:lambda1F1}
    \lambda_{\mathrm{1F-1}}
    =
    \frac{s\left[
    p^2-p\bigl(\sqrt{\Delta_{p,s}}+3s-3\bigr)
    -2(s-2)(s-1)
    \right]}
    {2(p-s)D_{p,s}}\,,
\end{equation}
where
\begin{align}
    \Delta_{p,s}
    &=p(p+6)-6ps+s^2+6s-7,\\
    D_{p,s}
    &=p^2+3(p-1)s-3p+s^2+2.
\end{align}
For $p=3$, $\Delta_{3,s}=(s-2)(s-10)$, so that a physical line exists only for $s\geq10$. At $s=10$, Eq.~\eqref{eq:lambda1F1} gives $\lambda=20/21$, where the four transition lines in Fig.~\ref{fig:PD3s} meet.

\textit{Asymptotic energy.—}
Gaussian integration by parts over the couplings gives the exact
finite-time energy density
\begin{equation}\label{eq:Et}
    E(t)
    =-\int_0^t\dd s\,
      f'(C(t,s))R(t,s).
\end{equation}
In the long-time limit, the integral separates into stationary and
aging contributions. Since the stationary correlations approach
$C=1$, Eq.~\eqref{eq:RTTI} gives
\begin{equation}
    E_{\mathrm{TTI}}
    =-f'(1)\int_0^\infty\dd\tau\,
      R_{\mathrm{TTI}}(\tau)
    =-\frac{f'(1)}{\sqrt{f''(1)}}.
\end{equation}
For the aging contribution, we use Eq.~\eqref{eq:RXC}. Changing the
integration variable from $s$ to $c=C(t,s)$ and using
$C(t,0)\to0$ as $t\to\infty$ yields
\begin{equation}
    E_{\mathrm{AG}}
    =-\int_0^1\dd c\,f'(c)X[c].
\end{equation}
Combining the stationary and aging contributions gives
Eq.~\eqref{eq:Einf} of the main text.

\end{document}